\documentclass[aps, prl, twocolumn, superscriptaddress, nofootinbib]{revtex4-1} 
\usepackage{amssymb, amsmath, amsthm}
\usepackage{xcolor}
\usepackage{graphicx}
\usepackage{hyperref}
\usepackage{comment}
 
\usepackage{footnote}
\usepackage{flushend}

\graphicspath{{Figures/}}

\newcommand{\ket}[1]{| #1 \rangle}
\newcommand{\bra}[1]{\langle #1 |}

\newcommand{\braket}[1]{\langle #1 \rangle}

\usepackage{adjustbox}
\makeatletter
\newcommand{\fitimage}[2][\@nil]{
	\begin{figure}
		\begin{adjustbox}{width=0.9\textwidth, totalheight=\textheight-2\baselineskip-2\baselineskip,keepaspectratio}
			\includegraphics{#2}
		\end{adjustbox}
		\def\tmp{#1}%
		\ifx\tmp\@nnil
		\else
		\caption{#1}
		\fi
	\end{figure}
}
\makeatother

\newcommand{\id}{\mathbf{1}}

\newcommand{\mytitle}{
Spin-statistics relation for quantum Hall states}

\begin{document}
 
\title{\mytitle}      

\author{Alberto Nardin}
\email{alberto.nardin@unitn.it}
\affiliation{INO-CNR BEC Center and Dipartimento di Fisica,  Universit\`a di Trento, 38123 Povo, Italy.}

\author{Eddy Ardonne}
\email{ardonne@fysik.su.se}
\affiliation{Department of Physics, Stockholm University, AlbaNova University Center, 106 91 Stockholm, Sweden}
 
\author{Leonardo Mazza}
\email{leonardo.mazza@universite-paris-saclay.fr}
\affiliation{Universit\'e Paris-Saclay, CNRS, LPTMS, 91405 Orsay, France}

\begin{abstract}
We prove a generic spin-statistics relation for the fractional quasiparticles that appear in abelian quantum Hall states on the disk.
The proof is based on an efficient way for computing the Berry phase acquired by a generic quasiparticle translated in the plane along a circular path, and on the crucial fact that once the gauge-invariant generator of rotations is projected onto a Landau level, it fractionalizes among the quasiparticles and the edge. 
Using these results we define a measurable quasiparticle fractional spin that satisfies the spin-statistics relation. 
As an application, we predict the value of the spin of the composite-fermion quasielectron proposed by Jain; our numerical simulations agree with that value. 
We also show that Laughlin's quasielectrons satisfy the spin-statistics relation, but carry the wrong spin to be the anti-anyons of Laughlin's quasiholes.
We continue by highlighting the fact that the statistical angle between two quasiparticles can be obtained by measuring the angular momentum whilst merging the two quasiparticles. Finally, we show that our arguments carry over to the non-abelian case by discussing explicitly the Moore-Read wavefunction.
\end{abstract}

\maketitle

\paragraph{\textbf{Introduction} ---}
The spin-statistics theorem is one of the pillars of our description of the world and classifies quantum particles into bosons and fermions according to their spin, integer or half-integer~\cite{Pauli_1940}.
It was early noted that in two spatial dimensions this relation is modified and intermediate statistics exist, called anyonic~\cite{Leinaas_1977, Wilczek_1982_B}. These objects too satisfy a generalised spin-statistics relation (SSR), and it is common nowadays to speak of fractional spin and statistics~\cite{Lerda_Book, Khare_Book}. This type of SSR, which we also consider, arises in a non-relativistic, non-field-theoretic context~\cite{Balachandran_1993,Preskill_2004}.
 
The quantum Hall effect (QHE)~\cite{goerbig2009quantum, tong2016lectures} is the prototypical setup where anyons have been studied, and several of their remarkable properties have also been experimentally observed~\cite{Nakamura_2019, bartolomei_2020}. 
Whereas the notion of fractional statistics has been early applied to the localised quasiparticles of the QHE~\cite{Arovas_1984, Halperin_1984, Nayak_2008, Feldman_2021}, the notion of spin has been more controversial.
The existence of a fractional spin satisfying a SSR has been established for setups defined on curved spaces thanks to the coupling to the curvature of the surface~\cite{Li_1992, Li_1993, Einarsson_1995, read2008quasiparticle, Gromov_2016, Trung_2022}.
The extension of this notion to planar surfaces has required more care and it is not completely settled yet~\cite{Sondhi_1992, Leinaas2002, Comparin_2022}.

In this letter we prove an SSR for the abelian quasiparticles of the QHE on a planar surface, for arbitrary filling fractions, directly from the microscopic Hamiltonian 
under the generic assumption that a QHE state satisfies the screening property.
It does not require the notion of curvature and identifies an observable spin that is an emergent collective property unrelated to the physical SU(2) spin.
Several applications are presented.
First, we study the quasielectron (QE) wavefunctions proposed by Jain~\cite{Jain_1989} and by Laughlin~\cite{Laughlin_1983} for the filling factor $\nu=1/M$.  
Second, we show how the fractional statistics affects the total angular momentum of the setup.
Thirdly, we discuss how our arguments carry over to the non-abelian case.
Finally, we remark on an intrinsic ambiguity in the definition of the spin.

\paragraph{\textbf{The QHE model} ---}

We consider a two-dimensional (2D) system of $N$ quantum particles with mass $m$ and charge $q>0$ traversed by a uniform and perpendicular magnetic field $\vec B = B \hat e_z$, $B>0$. The cyclotron frequency and the magnetic length read $\omega = qB/m$ and $\ell_B = \sqrt{\hbar c/ (qB)}$. 
We adopt the standard parametrization of the plane $z_j = x_j+iy_j = |z_j| e^{i \phi_j}$.

The Hamiltonian is:
\begin{equation}
 H_0 = \sum_{i=1}^N \left( \frac{\pi_{i,x}^2+ \pi_{i,y}^2}{2m} + v( |z_i|) \right)+ \sum_{i<j} V_\text{int}(| z_i - z_j|)
 \label{Eq:Ham:0}
\end{equation}
where $\pi_{i,a} = p_{i,a} - (q/c) A_{a}( z_i)$ and $v( |z|)$ is a central confining potential. We assume that the interaction potential $V_\text{int}(| z|)$ is rotationally invariant, as the Coulomb interaction relevant for electrons~\cite{Haldane_1985} and the contact interaction relevant for cold gases~\cite{Regnault_2003}. 
We also assume that the ground state of~\eqref{Eq:Ham:0} is not degenerate and realizes an incompressible QHE state characterised by screening: in the presence of perturbations which do not close the energy gap the particles will arrange in such a way that the density of the system is everywhere the same except in an exponentially localized region close to the defects; gentle modifications of the confinement potentials fall into this class of perturbations, so that
the specific form of $v( |z|)$ is not important if we are only interested in the bulk.

We assume the presence of $N_{qp}$
pinning potentials  located at positions $\mathbf s_\alpha$; using the complex-plane parametrisation $\eta_\alpha = s_{\alpha,x} + i s_{\alpha,y} = |\eta_\alpha| e^{i \theta_\alpha}$ we write: 
\begin{equation}
 H_1(\boldsymbol \eta)  = \sum_{\alpha=1}^{N_{qp}} \sum_{i=1}^N V_\alpha(|z_i - \eta_\alpha|),
\end{equation}
where
$\boldsymbol \eta$ is a shorthand for $\eta_1, \ldots, \eta_{N_{qp}}$.
Since the pinning potentials might be different, we keep the subscript $V_\alpha$; they are all assumed to be rotationally invariant.

The ground state of the model $H_{\boldsymbol \eta} = H_0 + H_1(\boldsymbol\eta)$ is $\ket{\Psi_{\boldsymbol \eta}}$; we assume that it is unique and that it localises $N_{qp}$ quasiparticles at $\eta_\alpha$. By virtue of screening, the density is everywhere the same as in the absence of pinning potentials, except close to the defects and at the boundary.
Since the pinning potentials can be different, the quasiparticles need not be of the same kind. The set of $\eta_\alpha$ is completely arbitrary and rotational invariance is generically broken; in our discussion, we will assume that they are always kept far from the boundary.
These assumptions imply that $\ket{\Psi_{\boldsymbol \eta}}$ can be a smooth function of $\boldsymbol \eta$.

\paragraph{\textbf{Quasiparticle self-rotations} ---}

\begin{figure}[t]
 	\begin{adjustbox}{width=0.5\textwidth, totalheight=\textheight-2\baselineskip,keepaspectratio,right}
 		\includegraphics[width=\columnwidth]{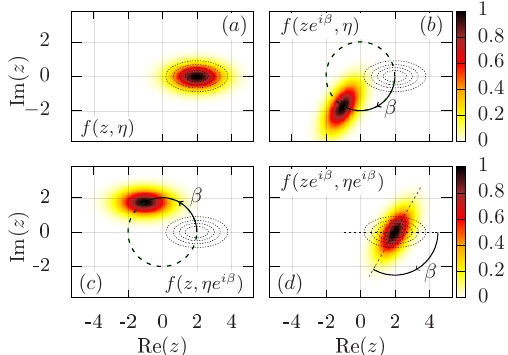}
 	\end{adjustbox}	
 \caption{Action of $\tilde L$. Panel $(a)$: contour plot in $z$ space of the function $f(z, \eta) = \exp[- \frac{1}{2}\Re(z-\eta)^2- 2 \Im(z-\eta)^2]$ for $\eta = 2$. Panel $(b)$: contour plot of $f(ze^{i\beta}, \eta)$ for $\beta = 2\pi/3$: with respect to $(a)$, the plot is translated and rotated. Panel $(c)$: contour plot of $f(z, \eta e^{i \beta})$: this time, the plot is only translated. Panel $(d)$: contour plot of $f(z e^{i \beta}, \eta e^{i \beta})$: the composition of the two is just a rotation and thus $\tilde L$ generates the self-rotations of the quasiparticles in the $z$ plane.}
 \label{Fig:Generator}
\end{figure}

We introduce the operator that is the sum of the particle angular momentum and of its quasi-particle generalisation measured in units of $\hbar$ (we use the symmetric gauge $\mathbf A = \frac 12 \mathbf B \wedge \mathbf r$),
\begin{equation}
	\tilde L = L_z + L_z'
	\; \; \text{with} \; \; L_z=-i\sum_{i=1}^{N}\frac{\partial}{\partial \phi_i},
		\; \; 
		L_z' = -i\sum_{\alpha=1}^{N_{qp}} \frac{\partial}{\partial\theta_\alpha}.
	\label{Eq:Tilde:L}
\end{equation}
We also define the group operator $U_\beta = e^{i \beta \tilde L}$ that is generated by~\eqref{Eq:Tilde:L}, with $\beta \in \mathbb R$.
The physical meaning of $U_\beta$ is best understood by considering its effect on a generic function $f(z, \eta)$, see Fig.~\ref{Fig:Generator}. 
Globally, $U_\beta$ is the composition of the two transformations, and represents the quasiparticle self-rotations over an angle $\beta$.  
 
%

Since the $\eta_\alpha$ are parameters, a gauge transformation $\ket{\Psi_{\boldsymbol \eta}} \to e^{i g(\boldsymbol \eta)} \ket{\Psi_{\boldsymbol \eta}}$ using an arbitrary smooth function of the parameters $g(\boldsymbol \eta)$ does not change the energy of the state.
Our goal is to show that it is always possible to use a gauge such that the ground-state is annihilated by $\tilde L$ and is thus invariant under the quasiparticle self-rotation operator $U_\beta$; this result is crucial for the proof of the SSR.

Let us first consider for simplicity the case of one quasiparticle, $N_{qp}=1$, so that $L'_z = -i \partial_{\theta}$ (we suppress the index $\alpha=1$ for brevity).
The Hamiltonian is explicitly invariant under the action of the group: $U_\beta H_\eta U_\beta^\dagger = H_\eta$.
With a quasiparticle at $\eta$, the ground state satisfies the Schr\"odinger equation $H_\eta \ket{\Psi_\eta} = E_\eta \ket{\Psi_\eta}$. However, $E_\eta$ can only depend on $|\eta|$, and not on $\theta$; thus, $\partial_\theta E_\eta = 0$. 
We conclude that
$
 H_\eta U_\beta \ket{\psi_\eta} = U_\beta E_\eta \ket{\Psi_\eta} = E_\eta  U_\beta \ket{\Psi_\eta},
$
namely that $U_\beta \ket{\Psi_\eta}$ is an eigenvector of $H_\eta$ with energy $E_\eta$.
If the ground state is unique, it must be an eigenvector of $U_\beta$ and of its generator $\tilde L$; we dub the eigenvalue of the latter $\ell_\eta$. 
For example, in the case of the normalised Laughlin state with a quasihole (QH), $\mathcal N(|\eta|)^{-1/2}\prod_i (z_i-\eta) \prod_{j<k} (z_j-z_k)^M e^{- \sum_i |z_i|^2/4 \ell_B^2}$, the eigenvalue $\ell_\eta = \frac{M}{2}N(N-1)+N $
is the degree of the polynomial in $z_i$ and $\eta$.

We now perform a gauge transformation that unwinds the generalised angular momentum $\ell_\eta$ moving along a trajectory at fixed $|\eta|$:
$
 \ket{\tilde \Psi_\eta} = e^{ i \tilde g(\eta)} \ket{\Psi_\eta} \label{Eq:Gauge:Trafo}
$
with $\tilde g(\eta) = - \int_0^\theta \ell_{|\eta|e^{i \theta'}} d \theta'$.
In the aforementioned case, the Laughlin state gets multiplied by the phase $(\eta/ \eta^*)^{- \frac{M}{4}N(N-1)- \frac{N}{2}}$.
Let us show that:
\begin{equation}
 \tilde L \ket{\tilde \Psi_\eta} = 0.\label{Eq:Kernel:Tilde:L}
\end{equation}
By definition, $\tilde L \ket{\tilde \Psi_\eta} = e^{i \tilde g(\eta)} \tilde L \ket{\Psi_\eta} + \left(L' e^{i \tilde g(\eta)} \right) \ket{\Psi_\eta}$.
The first term of the sum is $\ell_\eta \ket{\tilde \Psi_\eta}$, the second term is obtained by differentiating the exponential, and equals $-\ell_\eta \ket{\tilde \Psi_\eta}$.
This concludes the proof of the statement that one can find a gauge such that $\tilde L$ annihilates the ground state.
Note that for a state satisfying Eq.~\eqref{Eq:Kernel:Tilde:L}, it is also true that $U_\beta \ket{\tilde \Psi_{\eta}}= \ket{\tilde \Psi_{\eta}}$ for any angle $\beta$. Choosing $\beta = 2\pi$ we obtain that this state is single-valued in the $\eta $ coordinate because $U_{2\pi}\ket{\tilde \Psi_\eta}$ is also equal to $\ket{\tilde \Psi_{\eta e^{i 2 \pi}}}$.

This reasoning can be easily extended to the case of several quasiparticles. 
We can define a reference angle $\theta_0$ and express $\theta_\alpha = \theta_0 + \Delta \theta_{\alpha}$, treating the variables $\Delta \theta_{\alpha} = \theta_\alpha - \theta_0$ as independent from $\theta_0$. The operator $-i \partial_{\theta_0}$ generates the group $e^{i \beta \times (- i \partial_{\theta_0})}$ that modifies the quasiparticle polar angles as follows: $\theta_\alpha \to \theta_\alpha + \beta$, leaving the radial distance unchanged; thus: $\eta_\alpha \to \eta_\alpha e^{i \beta}$. This is exactly the action of $L'_z$, and thus we conclude that $L'_z = - i \partial_{\theta_0}$.
With arguments paralleling those for one quasiparticle, one can (i) show that 
$\tilde L \ket{\Psi_{\boldsymbol \eta}} = \ell_{\boldsymbol \eta} \ket{\Psi_{\boldsymbol \eta}}$, (ii) make the dependence on $\theta_0$, the $\Delta \theta_{\alpha}$ and the $|\eta_\alpha|$ explicit by writing $\ell_{\theta_0, \Delta \theta_\alpha, |\eta_\alpha|}$, and (iii) define $\ket{\tilde \Psi_{\boldsymbol \eta}} = e^{i g(\boldsymbol \eta)} \ket{\Psi_{\boldsymbol \eta}} $ with $ g(\boldsymbol \eta)=-  \int_0^{\theta_0} \ell_{\theta_0', \Delta \theta_\alpha, |\eta_\alpha|} d \theta_0'$, which is in the kernel of $\tilde L$.

\paragraph{\textbf{Berry phase for the translation of the quasiparticles along a circle} ---}
We now compute the Berry phase corresponding to the translation along a closed circular path of the $N_{qp}$ quasiparticle coordinates via $\theta_0 \to \theta_0+ 2\pi$ generated by $L'_z$ leaving all the $\Delta \theta_\alpha$ and $|\eta_\alpha|$ invariant. 
Using the fact that $\ket{\tilde \Psi_{\boldsymbol \eta}}$ is single-valued in $\boldsymbol \eta$, this Berry phase is $\gamma_{\boldsymbol{\eta}} = \int_0^{2\pi} \bra{\tilde\Psi_{\boldsymbol{\eta}}}i \partial _{\theta_0}\ket{\tilde\Psi_{\boldsymbol{\eta}}}d\theta_0$, where only the $\theta_0$ coordinate is changed in the state inside the integral.
Employing the definitions of $\tilde L$ and $L_z'$, and using \eqref{Eq:Kernel:Tilde:L}, we get
\begin{equation}
	\label{eq:BerryPhase}
	\gamma_{\boldsymbol{\eta}} = \int_0^{2\pi} \bra{\tilde\Psi_{\boldsymbol{\eta}}}L_z\ket{\tilde\Psi_{\boldsymbol{\eta}}}d\theta_0 
	 = \int_0^{2\pi} \bra{\Psi_{\boldsymbol{\eta}}}L_z\ket{\Psi_{\boldsymbol{\eta}}}d\theta_0. 
\end{equation}
The matrix element in the integral is manifestly gauge-independent, as the $L_z$ operator does not act on the $\boldsymbol \eta$; one can thus also use the original states. Finally, let us note that the integrand cannot be a function of $\theta_0$, and thus we have an even simpler expression: $\gamma_{\boldsymbol \eta}= 2 \pi \bra{\Psi_{\boldsymbol{\eta}}}L_z\ket{\Psi_{\boldsymbol{\eta}}}$.
This result was first established in Ref.~\cite{Umucalilar_2018} for the specific case of the Laughlin wavefunction and is here proved in full generality.  

Like any operator projected onto the lowest Landau level (LLL), the angular momentum $L_z$ is a function of the guiding-center operators~\cite{goerbig2009quantum} $R_{j,x} = x_j+ (\ell_B^2 /\hbar) \pi_{j,y} $ and
$R_{j,y} = y_j-(\ell_B^2 /\hbar) \pi_{j,x} $, with $[R_{j,x}, R_{j',y}] = - i \ell_B^2 \delta_{j,j'}$,  and it reads $L_z = \sum_j (R_j^2/\ell_B^2 - 1)/2$.
Written in this projected form, $L_z$ is the gauge-invariant generator of rotations, and it is just a function of the density of the gas $\rho_{\boldsymbol \eta}(z)$, which through the screening property can be split into a bulk contribution $\rho_b(z)$ (the state without quasiparticles), an edge contribution $\rho_e(z)$ (the difference at the edge with respect to the state without quasiparticles) and a quasiparticle contribution localised around the $\eta_\alpha$, $\rho_{qp, \boldsymbol \eta}(z)$. We split
the integrand into three parts:
$
	\bra{\Psi_{\boldsymbol{\eta}}}L_z\ket{\Psi_{\boldsymbol{\eta}}}  =L_b + L_e(N_{qp}) + L_{qp}(\boldsymbol{\eta});
$
as long as the quasiparticles are far from the edge, the screening property ensures that $L_e$ can only depend on their number (more precisely: on how many quasiparticles of each kind), but not on their positions; in fact, it also does not change when two of them are put close by or stacked on top of each other.

Notice that $L_b$ is an integer thanks to rotational invariance; 
therefore we disregard this contribution to the Berry phase \eqref{eq:BerryPhase}.
The only relevant information is contained in the remaining pieces, which indeed depend, directly or indirectly, on the quasiparticles, and this constitutes 
the first main result of the letter:
\begin{equation}
	\label{eq:Berry_noBulk}
	\gamma_{\boldsymbol{\eta}} = 2\pi \times \big(L_e(N_{qp}) + L_{qp}(\boldsymbol{\eta})\big).
\end{equation}
Compared to the direct computation of the integral, Eq.~\eqref{eq:Berry_noBulk} is simpler to evaluate.

Let us consider now the case of a single quasiparticle at $\eta$; on the basis of very general arguments, $\gamma_\eta$ should be the Aharonov-Bohm (AB) phase
$qQ\pi  |\eta|^2 B / (\hbar c)$, where
$Q$ is the charge of the quasiparticle in units of $q$.
Let us compare Eq.~\eqref{eq:Berry_noBulk} with this widely-accepted result.
In very general terms, the angular momentum of a rotationally-invariant quasiparticle $L_{qp}(\eta)=\int d^2r \left(r^2/2 \ell_B^2-1 \right) \rho_{qp,\eta}(r)$ can be split into an orbital part $ \frac{Q|\eta|^2}{2 \ell_B^2}$ 
and an intrinsic part 
\begin{equation}
	\label{eq:spinDef}
	J_{qp} = \int d^2r \left(r^2/2 \ell_B^2-1 \right) \rho_{qp,\eta=0}(r) = L_{qp}(0).
\end{equation}
It follows that
$\gamma_\eta =  \pi Q |\eta^2| / \ell_B^2 + 2 \pi(L_e(1)+ J_{1qp})$.
We recognise the AB phase, to which an apparently spurious contribution has been added; yet, we can show that it is an integer multiple of $2 \pi $, and thus inessential.
To show that $L_e(1)+ J_{1qp}$ is an integer, we consider a system with a QP {\em at its centre}, which is rotationally invariant, so its angular momentum $L_b + J_{1qp}+L_e(1)$ 
is an integer;
since $L_b\in\mathbb Z$, $J_{1qp}+L_e(1)$ is also an integer.
By the same logic $J_{nqp}+ L_e(n)\in \mathbb Z$ where $J_{nqp}$ is the spin of the rotationally symmetric QP obtained by fusing $n$ QPs together, stacking them on top of each other.
Very generically, the gauge-invariant generator of rotations fractionalizes between the bulk quasiparticles and the edge, implying that the spin is robust to local circularly-symmetric perturbations.
Before continuing, we mention a set of earlier works that have studied the properties of the second moment of the depletion density of fractional quasiparticles, which is shown to be related to the conformal dimension~\cite{Can_2015, Can_2016, Schine_2019}.

\paragraph{\textbf{Spin-statistics relation} ---}
We consider two identical quasiparticles  placed at opposite positions $\eta$ and $-\eta$ and far from each other and from the edge.
In order to compute the statistical parameter $\kappa$, we consider a double exchange, that gives a gauge-invariant expression and avoids any discussion on the identity of the pinning potentials~\cite{Arovas_1984}.
Accordingly, we study the difference between the Berry phase for exchanging two opposite particles and the single-particle AB phases~\cite{Kjonsberg_1999}:
\begin{equation}
	\kappa_{qp} = \frac{1}{2\pi} \left( \gamma_{\eta,-\eta} - 2 \gamma_\eta \right).
	\label{eq:TwiceLeinaasStatistics}
\end{equation}

Using Eq.~\eqref{eq:Berry_noBulk}, we write
$
	\kappa_{qp} =  L_e(2) + L_{qp}(\eta,-\eta) - 2 L_e(1) - 2L_{qp}(\eta).
$
As long as the QPs are well separated and since the liquid is screening, $\rho_{qp}(\eta,-\eta)=\rho_{qp,\eta}+\rho_{qp,-\eta}$ and thus $L_{qp}(\eta,-\eta) = 2L_{qp}(\eta)$; since $L_e(n)+J_{nqp} \in \mathbb Z$
we obtain the SSR:
\begin{equation}
 \kappa_{qp} = -J_{2qp} + 2 J_{1qp} \pmod 1.
 \label{Eq:SSR}
\end{equation}
This result allows us to identify the intrinsic angular momentum with the fractional spin associated to the fractional statistics, and constitutes the second main result of the letter. 
Interestingly, we have linked the statistics to a local property of the quasiparticles: if we assume screening, the fine details of the boundary do not matter, and one could probably prove~\eqref{Eq:SSR} without requiring that $v(|z|)$ is a central potential.

With similar arguments, the SSR can be extended to the situation where the two quasiparticles are different: calling $J_a$ and $J_b$ their spins, and $J_{ab}$ the spin of the composite quasiparticle obtained by stacking them at the same place, we obtain the mutual statistics parameter: $
 \kappa_{ab} = -J_{ab} + J_a + J_b \pmod 1.
$
In the theory of modular tensor categories (see see for instance \cite{kitaev06}), a relation of this type is called a \textit{ribbon identity}.
Moreover, the  fractionalisation property allows us to read the phase $\kappa$ directly at the edge; indeed, one easily obtains $\kappa_{qp} = L_e(2)-2 L_e(1)$ and $\kappa_{ab} = L_e(a,b) - L_e(a)-L_e(b)$.

\paragraph{\textbf{The spin of the QE} ---}
As a first application of our SSR~\eqref{Eq:SSR}, we consider the QE of the Laughlin state at filling $\nu = 1/M$. Numerical studies have highlighted that the composite-fermion wavefunction for the QE proposed by Jain~\cite{Jain_1989, Jeon_2003b} has the  correct statistical properties when the QE is braided with another QE ($\kappa_{qe} = 1/M$) or with a QH ($\kappa_{qe-qh} = -1/M$)~\cite{KJONSBERG1999705, Jeon_2003, Jeon_2004, Jeon_2005, Jeon_2010, Kjall_2018}.
Previous articles have already shown that LLL quasiparticles composed of $p$ stacked QHs fractionalise the angular momentum $J_p = - p^2 / (2M) + p/2$, and that these results are compatible via the SSR~\eqref{Eq:SSR} with a correct QH statistics $\kappa_{qh} = 1/M$~\cite{Comparin_2022, Trung_2022}.

\begin{table}[t]
 \begin{tabular}{|c||c|c|c|}
  \hline \hline
  & $\nu = \frac 12$ & $\nu = \frac 13$ & $\nu = \frac 14$ \\
  \hline \hline 
  $p = -1$ & $-\frac 34$ & $- \frac 23$ & $- \frac 58$ \\
  \hline
 \end{tabular} \hfill
 \begin{tabular}{|c||c|c|c|}
  \hline \hline
  & $\nu = \frac 12$ & $\nu = \frac 13$ & $\nu = \frac 14$ \\
  \hline \hline 
  $p = -2$ & $-2$ & $-\frac 53$ & $-\frac 32$\\
  \hline
 \end{tabular}  
 \caption{The spin $J_p $ of Jain's QE at filling factor $\nu$.}
\label{Table:Jain:QE}
\end{table}

\begin{figure}[t]
	\begin{adjustbox}{width=0.5\textwidth, totalheight=\textheight-2\baselineskip,keepaspectratio,right}
		\includegraphics[width=\columnwidth]{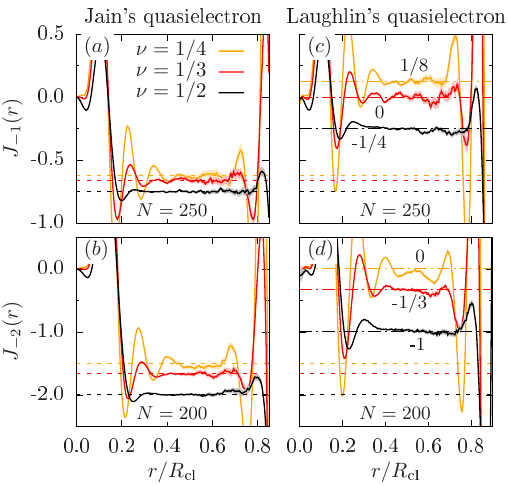}
	\end{adjustbox}	
 \caption{Calculation of the QE spin via the integral $J(r) = \int_0^r \left(\frac{|r'|^2}{2 \ell_B^2}-1 \right) \rho_{qp}(r) 2 \pi r' dr'$; the spin of Eq.~\eqref{eq:spinDef} coincides with the plateau appearing when $r$ is far from the center and the boundary; $R_{\rm cl}= \sqrt{2 N / \nu}$ is the classical radius of the droplet. 
 Panel $(a)$: the spin of a single Jain's QE for $\nu = 1/2$, $1/3$ and $1/4$. Panel $(b)$: the case of two stacked Jain's QEs. 
 Panels $(c)$ and $(d)$: the same for one and two Laughlin's QEs, respectively.  
 Theoretical predictions following from the SSR relation in Table~\ref{Table:Jain:QE} are marked with dashed lines and are only compatible with the spin of Jain's QE.
 Dashed-dotted lines in panels $(c)$ and $(d)$, together with their values, highlight the position of the spin plateau for Laughlin's QE.}
 \label{Fig:Jain:Spin}
\end{figure}

On the basis of these results and of the SSRs, it is easy to predict that Jain QE fractionalises the same spin $J_p$, with $p<0$ for QEs and $p>0$ for QHs.
We numerically verify this statement by performing a Monte-Carlo analysis of Jain's wavefunction with one QE $(p=-1)$ or two QEs $(p=-2)$~\cite{SM}.
Table~\ref{Table:Jain:QE} summarizes the expected values.
The results of our simulations are in Fig.~\ref{Fig:Jain:Spin}, panels $(a)$ and $(b)$, and they agree perfectly with our theory.
In the Supplemental Material~\cite{SM} we show the same results obtained with the Matrix-Product-State formulation~\cite{Zaletel_2012, Estienne_2013b, Hansson_2009, Hansson_2009b, Kjall_2018} using the Landau gauge. 
As we anticipated, our definition of quasiparticle spin is gauge invariant, and even if the two simulations are performed in different gauges (the symmetric and the Landau ones) the results coincide.  
Remarkably, this way of assessing the statistics of Jain's QE does not suffer from the undesired multi-particle position shift that needs to be taken into account in order to get the correct statistical phase~\cite{Jeon_2003, Kjall_2018}.

Concerning the QE wavefunction proposed in the original article by Laughlin~\cite{Laughlin_1983},  it was shown that it fractionalizes the correct charge, without making definitive statements about its braiding properties~\cite{Kjonsberg_1999, KJONSBERG1999705, Jeon_2003, Jeon_2004, Jeon_2010}.
The results of our numerical simulations are in Fig.~\ref{Fig:Jain:Spin}, panels $(c)$ and $(d)$. The plateau values are described by the spin $J'_p = -p^2 /(2M)+ p (2-M)/(2M)$, that gives the correct braiding phase for the Laughlin QEs, but that also shows that it is not the anti-anyon of the Laughlin's QH. 

\begin{figure}[t]
	\begin{adjustbox}{width=0.5\textwidth, totalheight=\textheight-2\baselineskip,keepaspectratio,right}
		\includegraphics[width=\columnwidth]{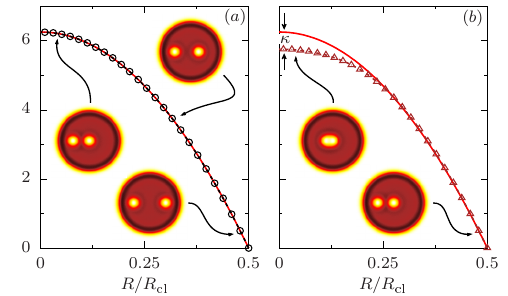}
	\end{adjustbox}
 \caption{Angular momentum $L(R_1, R_2)$ of a Laughlin state ($N=25$, $\nu = 1/2$) with two QHs at distances $R_1$ and $R_2$ from the center, computed with Monte Carlo techniques~\cite{Metropolis_1949, Hastings_1970}. 
 Panel $(a)$: displacement of the first QH; the angular momentum variation $L(R, R_0) - L(R_0,R_0)$ is plotted in black circles, and it is a quadratic function of $R$ that agrees with the theory prediction $- \epsilon (R^2 - R_0^2)$ (red line). 
 Panel $(b)$: displacement of the second QH; the variation $L(0,R)-L(0,R_0)$ is plotted in brown triangles and it is a quadratic function of $R$ only at large $R$; when 
 the QHs fuse a deviation sets in that equals $-\kappa$, the statistical parameter.}
 \label{Fig:AngularMomentum}
\end{figure}

\paragraph{\textbf{Angular-momentum of the gas} ---}
As a further application of the SSR, let us consider what happens when two QHs placed far apart are displaced radially in the sample. Let us call $L_0$ the angular momentum of the initial state with both quasiparticles at the same distance $R_0$ from the center. The first QH is then moved to the center: 
during this process the angular momentum increases and depends on the distance $R$ as  $L(R) = L_0 - \epsilon (R^2- R_0^2)$ with $\epsilon =qQ \pi B/(hc)$~\cite{SM}.
A gain in angular momentum of $\epsilon R_0^2$ is expected at the end of the process.
The same is now done with the second QH. Whereas also in this case the angular momentum increases, it does not attain the value $L_0 + 2 \epsilon R_0^2$ because when the two QH fuse, their total spin changes. In fact, the final value is $L_0 + 2 \epsilon R_0^2 - \kappa$.
We verify this result with numerical simulations reported in Fig.~\ref{Fig:AngularMomentum}.
This provides an experimental procedure for measuring the mutual statistics of two generic quasiparticles in a controllable quantum simulator of the QHE.

\paragraph{\textbf{The non-abelian case} ---}
Our arguments carry over to non-abelian QHE states (see Ref.~\cite{Macaluso_2019} for some earlier ideas). When considering a state with two QPs in a definite fusion channel, the ground state is actually unique (we restrict ourselves to the case without fusion multiplicities);
therefore, even the non-abelian case is covered, because the hypotheses of derivation of the SSR are uniqueness of the ground state, screening and rotational invariance.
The non-abelian nature shows up via the possibility that fusing two QPs can lead to different anyons, labeled by $c$.
There is a different SSR for each possibility,
$\kappa_{ab,c} = -J_c + J_a + J_b \pmod 1$.

As an example we discuss the SSR for the Moore-Read (MR) state~\cite{Moore_1991}.
We write the filling fraction of the state as $\nu = \frac{1}{q}$, where $q$ is even in the fermionic case and odd in the bosonic one.
The MR state is defined in terms of a chiral boson field $\varphi$ and the fields of the Ising conformal field theory~\cite{byb}.
This means that we should label the quasiholes by their Ising sector (i.e., $\mathbf 1$, $\sigma$ or $\psi$), and their charge.
The smallest charge quasihole has the labels $\left(\sigma,\frac{1}{2q}\right)$.
Because the fusion of two $\sigma$ fields has two possible outcomes, $\sigma\times\sigma = 1+ \psi$, the fusion of two quasiholes also leads to two possible results.
In particular, we have (the charge label is additive, as is the case for the Laughlin state)
\begin{equation}
\left(\sigma,\frac{1}{2q}\right) \times \left(\sigma,\frac{1}{2q}\right) = \left(\mathbf 1,\frac{1}{q}\right) + \left(\psi,\frac{1}{q}\right) \ .
\end{equation}
The first possible outcome $\left(\mathbf 1,\frac{1}{q}\right)$ is the quasihole one obtains by piercing the sample with an additional flux, i.e., the ``ordinary" Laughlin quasihole.
The second possible outcome ``contains" an additional neutral fermionic mode $\psi$.
Table~\ref{Table:MR:Spin} summarizes the expected value of the spin $J_a$ for each of the aforementioned quasiparticles~\cite{Bonderson_2011, SM}.   
We perform a Monte-Carlo sampling of the MR wavefunction with the different quasiparticles localised in the center of the system. The numerical calculations reported in Fig.~\ref{fig:MooreReadSpins} agree with the expected results~\cite{SM}.

\begin{table}[t]
 \begin{tabular}{|c|c|c|}
  \hline \hline
   $J_{\left( \mathbf 1, \frac 1q \right)}$ & $J_{\left( \sigma, \frac 1{2q} \right)}$ & $J_{\left(\psi,  \frac 1 q \right)}$ \\
  \hline \hline 
   $\frac 12$ & $\frac{1}{8q}+ \frac 3{16}$ & $0$ \\
  \hline
\end{tabular}
\caption{The spin $J_p $ of the MR quasiparticles.}
\label{Table:MR:Spin}
\end{table}

\paragraph{\textbf{Alternative spins} ---}

\begin{figure}[t]
\includegraphics[width=\columnwidth]{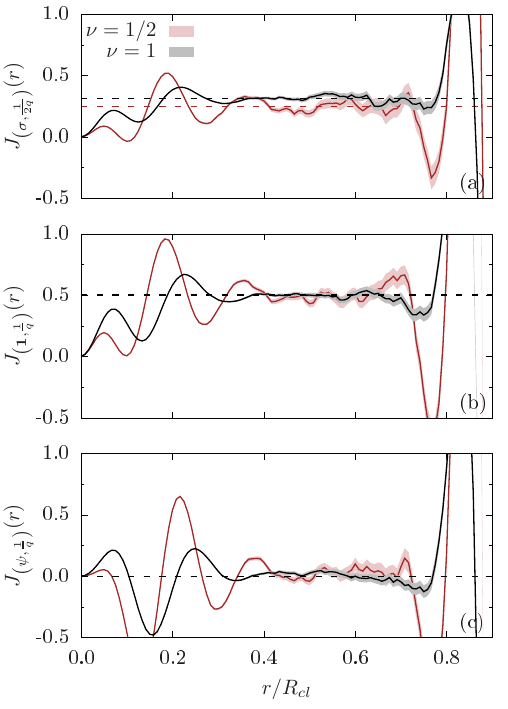} 
\caption{Comparison of the quasihole spins $J(r) = \int_0^r \left(\frac{{|r'|}^2}{2 \ell_B^2}-1\right) \left(\rho_{qp}(r')-\rho_{MR}(r')\right) 2 \pi r' dr'$, where $\rho_{qp}(r)$ is the QH density and $\rho_{MR}(r)$ the background Moore-Read density, for the different Moore-Read quasiholes: (a) the $\left(\sigma,\frac{1}{2q}\right)$, (b) the $\left(\mathbf{1},\frac{1}{q}\right)$ and (c) the $\left(\psi,\frac{1}{q}\right)$, for the bosonic filling $\nu=1$ (corresponding to $q=1$) and the fermionic $\nu=\frac{1}{2}$ ($q=2$).
$R_{\rm cl}= \sqrt{2 N / \nu}$ is the classical radius of the droplet and the number of particles is $N=200$. } 
\label{fig:MooreReadSpins}
\end{figure}

Our definition of spin follows directly from the physical angular momentum 
$L=\mathcal{L}_R+\mathcal{L}_\pi$, where
$\mathcal{L}_R=(R_x^2+R_y^2)/2l_B^2$
and
$\mathcal{L}_\pi=-(\pi_x^2+\pi_y^2)l_B^2/2\hbar^2$;
it is manifestly gauge-invariant and is the generator of 2D rotations, because it satisfies $[L,R_j]=i\epsilon_{jk}R_k$ and $[L,\pi_j]=i\epsilon_{jk}\pi_k$, with $\epsilon_{jk}$ the Levi-Civita tensor.
The definition is ambiguous: any operator $\mathcal L_c = L + c$, with $c$ a c-number, has indeed the correct commutation properties:
this is a peculiarity of U(1) rotations in two-dimensional physics, as SU(2) ones do not leave room for such ambiguity.
We conclude that any operator $\mathcal L_c$ defines a correct quasiparticle spin $J_p(c)$~\cite{Trung_2022}. In very general terms, $J_p$ is composed of a part proportional to $p^2$ that determines the anyonic statistics~\cite{Thouless_1985}, and of a part proportional to $p$ that does not affect $\kappa_{qp}$, see~\eqref{Eq:SSR}.
It is not difficult to prove that $c$ can only appear in the prefactor multiplying $p$, as it is linear in the quasiparticle density.
We consider this as an essential ambiguity that cannot be resolved, although different choices may have different physical meanings.

\paragraph{\textbf{Conclusions} ---}

We have presented a SSR for the abelian quasiparticles of the QHE on planar surfaces derived from very mild assumptions.
We have shown that the quasiparticles fractionalise the gauge-invariant generator of rotations and that this quantity can be used to define a measurable spin. The fractional statistical properties of the quasiparticles follow from that.
Our results carry over to non-abelian quantum Hall states.

\paragraph{\textbf{Acknowledgements} --- }
We acknowledge discussions with I.~Carusotto, T.~Comparin, B.~Estienne, H.~Hansson, M.~Hermanns, A.~Polychronakos and N.~Regnault.
A.N.~thanks Universit\'e Paris-Saclay and LPTMS for warm hospitality.
L.M.~has been supported by LabEx PALM (ANR-10-LABX-0039-PALM).

\bibliography{SpinStatistics.bib}

\setcounter{secnumdepth}{5}
\addtocontents{toc}{\protect\setcounter{tocdepth}{0}}
\renewcommand{\theequation}{S.\arabic{equation}}
\renewcommand{\thefigure}{S\arabic{figure}}
\renewcommand{\thesection}{S\arabic{section}}
\setcounter{equation}{0}
\setcounter{figure}{0}


\newpage
\clearpage
\onecolumngrid

\begin{center}
\begin{Large}
\textbf{Supplementary material for}

\textbf{\mytitle}
\end{Large}

\vspace{0.25cm}

	Alberto Nardin,$^{1}$ Eddy Ardonne,$^{2}$ and Leonardo Mazza$^{3}$

	\vspace{0.25cm}	

	$^{1}$\textit{INO-CNR BEC Center and Dipartimento di Fisica,  Universit\`a di Trento, 38123 Povo, Italy.}
	
	$^{2}$\textit{Department of Physics, Stockholm University, AlbaNova University Center, 106 91 Stockholm, Sweden}
	
	$^{3}$\textit{Universit\'e Paris-Saclay, CNRS, LPTMS, 91405 Orsay, France}

	\today
	
	\end{center}

	\onecolumngrid
	
	\setcounter{secnumdepth}{4}

	\section{The spin of the quasielectron}
	We tested the spin of two paradigmatic quasielectron (QE) wavefunctions: Laughlin's~\cite{Laughlin_1983} and Jain's~\cite{Jain_1989}, both for single and double QE.
	The spin of lowest Landau level projected wavefunctions is computed according to~\cite{Comparin_2022}
	\begin{equation}
		\label{eq:sm_spin_def}
		J_{qe}(R) = 2\pi\int_0^R r\,dr\, \left(\frac{r^2}{2}-1\right) (\rho_{qe}(r)-\rho(r)),
	\end{equation}
	$\rho_{qe}(r)$ being the density of a QE state placed at the origin and $\rho(r)$ the background density of the fractional quantum Hall state hosting the QE excitation.
	When $1 \ll r \ll R_{cl}$, $J_{qe}(R)$ has a plateau at the spin value.	
	The knowledge of the analytical form of the wavefunctions allows us to compute the spin by Monte Carlo sampling~\cite{Metropolis_1949} the spin integral \eqref{eq:sm_spin_def}.
	
\begin{figure}[b]
		\begin{adjustbox}{width=1\textwidth, totalheight=\textheight-2\baselineskip,keepaspectratio,right}
			\includegraphics[width=\columnwidth]{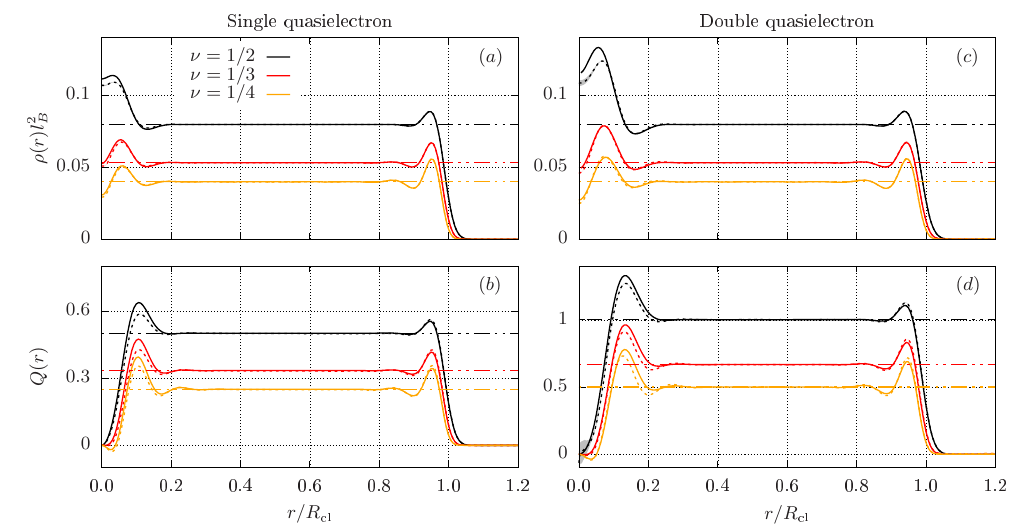}
		\end{adjustbox}	
		\caption{Comparison of Jain's and Laughlin's QE densities $\rho_{qp}(r)$ and charges $Q(r) = \int_0^r \left(\rho_{qp}(r')-\rho_{L}(r')\right) r' dr'$, where $\rho_{qp}(r)$ is the QE density and $\rho_{L}(r)$ the background Laughlin density. The charge $Q$ of the QE coincides with the plateau appearing when $r$ is far from the center and the boundary; $R_{\rm cl}= \sqrt{2 N / \nu}$ is the classical radius of the droplet. 	
		Panel $(a)$ (/$(c)$): comparison between the densities $\rho_{qp}(r)$ of a single (/double) Jain's QE (full lines)  compared to that of Laughlin's QEs (dashed lines), for $\nu = 1/2$, $1/3$ and $1/4$. Horizontal dashed-dotted lines represent the bulk Laughlin state density $\rho_b=\nu/2\pi l_B^2$.		
		Panel $(b)$ (/$(d)$): comparison between the QE charge $Q(r)$ of a single (/double) Jain's QE (full lines) compared to that of Laughlin's QEs (dashed lines). Horizontal dashed-dotted lines represent the charge of Laughlin's quasiparticles, $Q=\nu$.}
		\label{fig:ChargeAndDensity}
	\end{figure}

	In the following subsections, some information on the numerics is given.
	The results for the spin \eqref{eq:sm_spin_def} are shown in the main text; here in Fig. \ref{fig:ChargeAndDensity} we complement by showing the density of the states at filling fraction $\nu=1/2$ and the excess charge with respect to the bulk Laughlin liquid.

	\subsection{Single Jain's quasielectron}
	Jain's composite fermion approach to the fractional quantum Hall states suggests~\cite{Jain_1989}
	\begin{equation}
		\Psi_{JQE} = \hat{P}_\text{LLL} 
		\left|
		{
			\begin{array}{cccc}
				z_0^* & z_1^* & z_2^* & \hdots \\
				1 & 1 & 1 & \hdots \\
				z_0 & z_1 & z_2 & \hdots \\
				\vdots & \vdots & \vdots & \ddots \\
			\end{array}
		}
		\right| 
		\prod_{i<j}(z_i-z_j)^{m-1}
	\end{equation}
	as a candidate wavefunction for the QE on top of a Laughlin state at filling $\nu=\frac{1}{m}$. Here and in the following, Gaussian factors will be left implicit. 
	Carrying out standard projection onto the lowest Landau level~\cite{Girvin_1984} gives (apart for constant proportionality factors) 
	\begin{equation}
		\label{eq:sm_projected_jain_qe}
		\begin{split}
			\Psi_{JQE} 
			& = \sum_i \left(\frac{1}{\prod_{l\neq i} z_l-z_i}\sum_{j\neq i}\frac{1}{z_i-z_j}\right)\prod_{i<j}(z_i-z_j)^{m}
		\end{split}
	\end{equation}
	which has already been shown to carry the correct fractional charge~\cite{KJONSBERG1999705} and having the correct exchange statistics~\cite{Jeon_2003, Kjall_2018}.

	\subsection{Double Jain's quasielectron}
	Jain's composite fermion approach suggests the following wavefunction for a doubly charged QE at the centre of a circularly symmetric droplet
	\begin{equation}
		\Psi_{J2QE} = \hat{P}_\text{LLL} 
		\left|
		{
			\begin{array}{cccc}
				{z_0^*}^2 & {z_1^*}^2 & {z_2^*}^2 & \hdots\\
				z_0^* & z_1^* & z_2^* & \hdots \\
				1 & 1 & 1 & \hdots \\
				z_0 & z_1 & z_2 & \hdots \\
				\vdots & \vdots & \vdots & \ddots \\
			\end{array}
		}
		\right| 
		\prod_{i<j}(z_i-z_j)^{m-1}.
	\end{equation}
	Notice that this wavefunction is not the most energetically-favourable double quasielectron state~\cite{Jeon_2003b}, which is realized by promoting two composite fermions to their first Landau level.
	We study the composite fermion antiparticle of the double-quasihole state wavefunction because it has the same angular momentum as the Laughlin's double quasielectron state and therefore the comparison is more direct~\cite{Jeon_2005}.	
	\newline We use standard lowest Landau level projection, although different inequivalent projection methods~\cite{JainKamilla_IntJModPhysB_1997} have been proposed.
	After some tedious algebra (and again dropping constant proportionality factors) we find
	\begin{equation}
		\label{eq:sm_projected_jain_dqe}
		\begin{split}
			\Psi_{J2QE} 
			& = \sum_{i\neq j}\left( \frac{(z_i-z_j)\Gamma_{ij}}{\prod_{k\neq i}(z_k-z_i)\prod_{k\neq j}(z_k-z_j)}\right)\prod_{i<j}(z_i-z_j)^{m}
		\end{split}
	\end{equation}
	where 
	\begin{equation}
		\Gamma_{ij} = (m-1)^2A_i^2A_j-(m-1)B_iA_j+\frac{2(m-1)A_i}{(z_i-z_j)^2}-\frac{2}{(z_i-z_j)^{3}}
	\end{equation}
	and
	\begin{equation}
		\begin{cases}
			A_i=\sum_{j\neq i}\frac{1}{z_i-z_j}\\
			B_i=\sum_{j\neq i}\frac{1}{(z_i-z_j)^2}.
		\end{cases}
	\end{equation}
		
	\subsection{Single Laughlin's quasielectron}\label{par:1lqe}
	Laughlin proposed a QE wavefunction by generalizing his successful quasihole (QH) wavefunction~\cite{Laughlin_1983}
	\begin{equation}
		\label{eq:LaughlinsQE}
		\psi_{LQE} = \left(\prod_i 2\frac{\partial}{\partial z_i}\right)\prod_{i<j}(z_i-z_j)^{m},
	\end{equation}
	which however - unlike the QH counterpart - is not easy to deal with from the computational point of view, due to the $N$-th order derivative term. 
	We are interested in computing expectation values of local, single-particle observables $\hat{O}=\sum_i \hat{o}_i$
	\begin{equation}
		\label{eq:sm_local_observable}
		\braket{\hat{O}} = \int \mathcal{D}z\, \psi_{LQE}^* O(z, z^*) \psi_{LQE}
	\end{equation}
	where $\mathcal{D}z=\prod_i d^2z_i$ and $z$ is a shorthand for all the particles' coordinates. 
	To simplify the expressions we assume $O(z,z^*)=O(|z|^2)$. 
	By performing integration by parts $2N$ times we then get
	\begin{equation}
		\label{eq:sm_LaughlinQE_observable}
		\braket{\hat{O}} = \frac{1}{Z}\int \mathcal{D}z\, \Biggl|\prod_{i<j}(z_i-z_j)^{m}\Biggr|^2 \prod_i(|z_i|^2-2) \sum_i A_i(|z_i|^2)
	\end{equation}
	where $A_i$ is related to the observable $o_i$ through
	\begin{equation}
		A(r^2) = o(r^2)-4 \frac{r^2-1}{r^2-2} \frac{\partial o(r^2)}{\partial r^2}+4 \frac{r^2}{r^2-2} \frac{\partial^2 o(r^2)}{\partial (r^2)^2}
	\end{equation}
	but crucially involves its derivatives.
	The normalization factor $Z$ can be found by looking at $\hat{O}=\mathbb{I}.$
		
	As an example, we expand here the expressions for the observable being the charge up to radius $R$
	\begin{equation}
		o(r^2) = \theta(R^2-r^2).
	\end{equation}
	where $\theta$ is the step function.
	The spin case \eqref{eq:sm_spin_def} is perfectly analogous but the expressions are more lengthy because of the $r^2/2-1$ factor multiplying the step function.
	The integrals involve derivatives of the $\delta$ function. It is convenient to take these out of the integrals and rearrange \eqref{eq:sm_LaughlinQE_observable} in the following form
	\begin{equation}
		Q(R) = I_0(R^2) + 4 I_1(R^2) + 4 \frac{\partial I_2(R^2)}{\partial R^2}
	\end{equation}	
	where
	\begin{equation}
		\begin{cases}
			I_0(R^2) = \frac{1}{Z} \int\mathcal{D}z\, \left|\prod_{i<j}(z_i-z_j)^{m}\right|^2 \prod_i(|z_i|^2-2)\,\sum_i \theta(R^2-r_i^2)
			\\
			I_1(R^2) = \frac{1}{Z} \int\mathcal{D}z\, \left|\prod_{i<j}(z_i-z_j)^{m}\right|^2 \prod_i(|z_i|^2-2)\,\sum_i\delta(R^2-r_i^2) \frac{r_i^2-1}{r_i^2-2}
			\\
			I_2(R^2) = \frac{1}{Z} \int\mathcal{D}z\, \left|\prod_{i<j}(z_i-z_j)^{m}\right|^2 \prod_i(|z_i|^2-2)\,\sum_i \delta(R^2-r_i^2)	\frac{r_i^2}{r_i^2-2}.				
		\end{cases}
	\end{equation}
	Taking derivatives of noisy observables is tricky; to circumvent the problem we Fourier transform the relevant quantities
	\begin{equation}
		\tilde{I}(k) = \int_0^\infty I(R) J_0(k R) R\,dR
	\end{equation}
	where $J_0$ is the order $0$ Bessel function of the first kind.
	We then filter out the ``high-wavevector" noise superimposed to the ``low-wavevector" signal $\tilde{I}_c(k) = c(k)\tilde{I}(k)$, with some suitably chosen cut-off function $c(k)$, and invert the transform
	\begin{equation}
		\frac{\partial^n I_c(R)}{\partial (R^2)^n} = \int_0^\infty \tilde{I}_c(k) \frac{\partial^n J_0(k R)}{\partial (R^2)^n} k\,dk.
	\end{equation}
	Derivatives of $J_0$ can be expressed in closed compact form in terms of the $_0F_1$ hypergeometric function, thus avoiding the computation of finite differences.
		
	\subsection{Double Laughlin's quasielectron}\label{par:2lqe}
	A doubly charged Laughlin's QE can be placed at the origin as~\cite{Jeon_2003b}
	\begin{equation}
		\psi_{L2QE} = \left(\prod_i 2\frac{\partial}{\partial z_i}\right)^2\prod_{i<j}(z_i-z_j)^{m}.
	\end{equation}	
	Again, carrying out the derivatives explicitly seems not to be feasible, however it is simpler to look at local observables as \eqref{eq:sm_local_observable}. Repeated integration by parts yields
	\begin{equation}
	\label{eq:sm_Laughlin2QE_observable}
	\braket{\hat{O}} = \frac{1}{Z}\int \mathcal{D}z\, \Biggl|\prod_{i<j}(z_i-z_j)^{m}\Biggr|^2 \prod_i(8-8|z_i|^2+|z_i|^4) \sum_i A_i(|z_i|^2)
	\end{equation}
	with
	\begin{equation}
		\begin{split}
		A(r^2) = o(r^2)&-8 \frac{4-6r^2+r^4}{8-8r^2+r^4} \frac{\partial o(r^2)}{\partial r^2}+8 \frac{4-12r^2+3r^4}{8-8r^2+r^4}  \frac{\partial^2 o(r^2)}{\partial (r^2)^2}-\\&-32 \frac{-2r^2+r^4}{8-8r^2+r^4}  \frac{\partial^3 o(r^2)}{\partial (r^2)^3}+16\frac{r^4}{8-8r^2+r^4}  \frac{\partial^4 o(r^2)}{\partial (r^2)^4}.
		\end{split}
	\end{equation}
	Once again, the expressions for the observables involve derivatives of the observable itself; whenever dealing with derivatives of the delta function, we adopted the procedure outlined in the previous paragraph.
	
	\section{Angular momentum of the gas}
	We here discuss a bit more extensively the results presented in Fig. 3 of the main text. We considered a Laughlin state with two quasi-holes
	\begin{equation}
		\label{eq:multiple_laughlin_qh}
		\psi(\boldsymbol{\eta}_1,\boldsymbol{\eta}_2) = \prod_i(z_i-\boldsymbol{\eta}_1)\,\prod_i(z_i-\boldsymbol{\eta}_2)\,\prod_{i<j}(z_i-z_j)^m
	\end{equation}
	initially at diametrically opposite positions, $\eta_2=-\eta_1=R_0$.
	As long as the two quasi-holes are well within the bulk of the systems and far away from each other, the angular momentum of the state can be decomposed as
	\begin{equation}
		\label{eq:gasAngularMomentum}
		L(\boldsymbol{\eta}_1,\boldsymbol{\eta}_2) = L_b+L_e(2)+L_{qp}(\boldsymbol{\eta}_1)+L_{qp}(\boldsymbol{\eta}_2).
	\end{equation}
	We decompose the quasi-hole angular momenta exploiting the fact that their density profile is  circularly-symmetric and depends only on $\mathbf{r}$, $\boldsymbol{\eta}$ through $|\mathbf{r}-\boldsymbol{\eta}|$
	\begin{equation}
		L_{qp}(\eta_1) = \int d^2r \left(\frac{r^2}{2}-1\right)\rho_{qh}(|\mathbf{r}-\boldsymbol{\eta}_1|) = J_1 + \frac{Q_1 |\boldsymbol{\eta}_1|^2}{2}
	\end{equation}
	where $J_1$ is the single quasi-hole spin, while $Q_1=\int d^2r \rho_{qh}(r)$ its charge. As one quasi-hole is moved through the quantum Hall fluid at time $t_0$ and brought towards the centre at time $t_1$, the angular momentum variation reads 
	\begin{equation}
		\Delta L(\boldsymbol{\eta}_1,\boldsymbol{\eta}_2) = \frac{Q_1}{2}\,\, \left(|\boldsymbol{\eta}_1(t)|^2 - |\boldsymbol{\eta}_1(t_0)|^2\right)
	\end{equation}
	provided the two quasi-holes are far away from one other. 
	
	When also the second quasi-hole is brought towards the center
	a naive use of Eq.~\eqref{eq:gasAngularMomentum} would predict the same angular momentum variation, with final angular momentum $\widetilde L(0,0)=L_b+L_e(2)+2J_1$. 
	This is not correct though, since as soon as the two quasi-holes start fusing the assumption of them being far away from each other ceases to be valid. 
	The correct $\eta_1=\eta_2=0$ limit reads $L(0,0) = L_b+L_e(2)+J_2$, so the difference between the parabolic behaviour and the correct one reads $L(0,0)-\widetilde L(0,0)=J_2-2J_1=-\kappa$, the statistics parameter.
	
	\section{Monte-Carlo sampling}
	\label{Section:MCSampling}
	The numerical results presented in the main text and those in Fig. \ref{fig:ChargeAndDensity} in this Supplemental Materials are the results of Monte Carlo simulation; in particular, we used standard Metropolis-Hastings Monte Carlo algorithm\cite{Metropolis_1949,Hastings_1970} to sample configurations from Jain's \eqref{eq:sm_projected_jain_qe} and Laughlin's \eqref{eq:LaughlinsQE} wavefunctions (in the latter case, according to the methods described in the previous paragraphs \ref{par:1lqe} and \ref{par:2lqe}), as well as the quasi-hole state Eq.~\eqref{eq:multiple_laughlin_qh}.
	Observable expectation values $\braket{\hat{O}}$ have been obtained by collecting the results of multiple Monte-Carlo runs executed in parallel on a single GPU, for a total of $\approx 10^{10}\div10^{11}$ configurations, which were split into $M\simeq100$ independent realizations, $O_i$; the average and statistical errors are computed in the usual way, $\braket{\hat{O}}\simeq \frac{1}{M}\sum_i O_i$ and $\sigma_O^2\simeq \frac{1}{M(M-1)}\sum_i(O_i-\braket{\hat O})^2$. 
	The massive parallelism of the GPU allowed for relatively quick simulations (longest simulations taking up to some days), taking down the error-bars on Eq.~\eqref{eq:sm_spin_def}.
		
	Finally, let us comment on the system size we chose. We mainly focused on large systems of $N\gtrsim200$ when looking at the spin Eq.~\eqref{eq:sm_spin_def} because the density oscillations at both the quasiparticle “edge” and at the system's one grow larger with the reciprocal of the filling fraction, $m=1/\nu$; it can indeed be seen in Fig. 2(a) of the main text that the plateau is much more pronounced at $\nu=1/2$ than it is at $\nu=1/4$.	
	
	Numerical data can be shared upon reasonable request.

\section{The MPS formulation on the cylinder}

In the main text, the spin of the single and double QEs was obtained for the disk geometry.
Here, we report results on the cylinder, using the MPS approach, for the $\nu=\frac{1}{M}$ fermionic Laughlin state.
We refer to \cite{Zaletel_2012, Estienne_2013b} for more information on the MPS formulation of quantum Hall states in general.
The MPS formulation of a single Jain QE was given in detail in~\cite{Kjall_2018}.
Here, we provide the MPS matrices for a QH of size $p$, i.e., a QH with charge $\frac{p}{M}$, as well
as for QEs with various sizes, i.e., the cases with $p$ negative.

As was discussed in detail in~\cite{Kjall_2018}, if one wants to be able to consider a state with
several QHs and QEs, it is necessary to introduce {\em two} chiral boson fields $\varphi$ and $\tilde\varphi$,
in order that the various operators have the correct statistics with one another.
However, in the case that one is interested in simulating only one QE (which can be a single QE of arbitrary size),
without any QHs, a single chiral boson field suffices.

The CFT operators for an electron, a size $p>0$ QH and a the modified electron operator for the size $p<0$ QE (on the disk) are given by
(in terms of the single chiral boson field $\varphi(z)$)
\begin{align}
V_{\rm el} (z) &= : e^{i \varphi (z) \sqrt{M}} : \ , &
V_{\rm qh,p} (\eta) &= : e^{i \varphi (\eta) p/\sqrt{M}} : \ , &
\tilde{V}_{\rm el,p} (z) &= \partial_z^{|p|}: e^{i \varphi (z) (M-|p|)/\sqrt{M}} : \ ,
\end{align}
where we note that $p<0$ in the QE case. In addition, one has the following constraint on the size of the QE, $0< |p| < M$. 

Without going into the details, we here state the MPS matrices for an empty site, a site occupied by an electron,
the matrix for a size $p$ QH, as well as the modified electron operators corresponding to a size $p$ QE.

\subsection{The matrices of the empty sites and ordinary electron operator}

The circumference of the cylinder is denoted by $L$.
Using the notation of~\cite{Kjall_2018}, the MPS matrix for an empty site is given by
\begin{equation}
B^{[0]} = \delta_{Q',Q-1}\delta_{P',P}\delta_{\mu',\mu}
e^{-\bigl(\frac{2\pi}{L}\bigr)^2 \bigl( \frac{(Q')^2}{2M}+\frac{Q'}{2M}+P'\bigr)}
\ .
\end{equation}
The matrix for an site occupied by an electron is given by
\begin{equation}
B^{[1]} = \delta_{Q',Q+M-1}\delta_{P',P-Q} A^{\sqrt{M}}_{\mu',\mu}
e^{-\bigl(\frac{2\pi}{L}\bigr)^2 \bigl( \frac{(Q')^2}{2M}+\frac{Q'}{2M}+P'\bigr)}
\ ,
\end{equation}
where $A^{\beta}_{\mu',\mu}$ is given by
\begin{align}
A^{\beta}_{\mu',\mu} &=
\prod_{j=1}^{\infty} \sum_{s=0}^{m'_j} \sum_{r=0}^{m_j}
\delta_{m_j-r,m'_j-s}\frac{(-1)^r}{\sqrt{r! s!}} \Bigl(\frac{\beta}{\sqrt{j}}\Bigr)^{r+s} \sqrt{\binom{m'_j}{s}\binom{m_j}{r}} \ .
\label{eq:Avals}
\end{align}

\subsection{The matrices for the size $p$ QH operator}

We consider the matrix for the size $p$ QH operator, where it is assumed that we only consider states with
one QH (which can be of arbitrary size $p > 0$).
The operator is inserted between orbitals $l-1$ and $l$.
We denote this operator as $H_{l,p}(\eta)$, where $\eta$ is the position of the QH on the cylinder.
We find
\begin{align}
H_{l,p}(\eta) =
(-1)^{\frac{p}{M}(Q+l)}\delta_{Q',Q+p} A^{\frac{p}{\sqrt{M}}}_{\mu',\mu} \\
e^{-\bigl(\frac{2\pi}{L}\bigr)^2 (\tilde{\tau}_\eta-l)(\frac{Q^2-Q'^2+Q-Q'}{2M}+P-P')}
e^{-\bigl(\frac{2\pi}{L}\bigr)(i x_\eta)(P'-P+\frac{1}{M}(p Q - \tilde\tau_\eta))} \ .
\end{align}

\subsection{The matrices for the modified electron for the size $p$ QE operator}

As explained in detail in~\cite{Kjall_2018}, a QE is created by modifying an elektron operator, to create a
delocalized (angular momentum eigenstate, with angular momentum $k$) QE.
The latter are used to create a localized QE at $\xi$, by means of applying a localizing kernel, weighing the
angular moment QEs.
Here, we give the result for the modified electron operator, for the orbital $l$, that creates the localized QE
of size $p$ at $\xi$,
\begin{equation}
E_{l,p} (\xi) = e^{-\frac{\tau_\xi^2}{2 M}} \sum_k
e^{-\bigl( \frac{2\pi}{L}\bigr)^2 M k^2}
e^{\frac{2\pi}{L}k (ix_\xi+\tau_\xi)} E_{k,l,p} \ .
\end{equation}
Here, $E_{k,l,p}$ is given by
\begin{equation}
E_{k,l,p} =
(-1)^{|p|\frac{Q+l}{M}}
F
\delta_{Q',Q+M-|p|-1} \delta_{P',P-Q-k+|p|+\frac{|p|}{M}(Q+l)}
A^{\frac{q-|p|}{\sqrt{M}}}_{\mu',\mu}
e^{
-\bigl(\frac{2\pi}{L}\bigr)^2 (\frac{Q'^2}{2M}+\frac{Q'}{2M}+P')
}
e^{
\bigl(\frac{2\pi}{L}\bigr)^2 (l k + \frac{l^2 |p| + l |p|^2}{2M})
}
\ ,
\end{equation}
where the factor $F$ (which takes care of the derivative in the modified electron operator, and ensures that we have proper angular momentum states) is given by
\begin{equation}
F = \sum_{s=0}^{|p|} (-1)^s \binom{|p|}{s}
\Bigl(
\prod_{r_1=0}^{s-1} \frac{k-r_1}{|p| N_e -r_1} 
\Bigr)
\Bigl(
\prod_{r_2=s+1}^{|p|} (l-k+r_2) 
\Bigr)
\Bigl(
\prod_{r_3=1}^{s} (M(N_e-1)-|p|N_e+r_3) 
\Bigr) \ ,
\end{equation}
where the products are defined to be one if the range of the product is empty. This happens for $s=0$, $s=|p|$ and $s=0$ for the
first, second and third product in the sum, respectively.

\paragraph{\textbf{Results from the MPS formulation} ---}

We used the MPS formulation for the QH and QE states, to obtain the density profiles, the access charge $Q(r)$ and the spin $J(r)$ of the excitations.
In all cases, we used the following parameters for the MPS.
The number of electrons $N_e = 100$; the circumference of the cylinder $L=18 \ell_B$. The maximum value of the momentum for the non-zero modes
$p_{\rm max} = 12$.
In all cases, the excitations were placed in the middle of the cylinder.

\begin{figure}
\includegraphics[width=\textwidth]{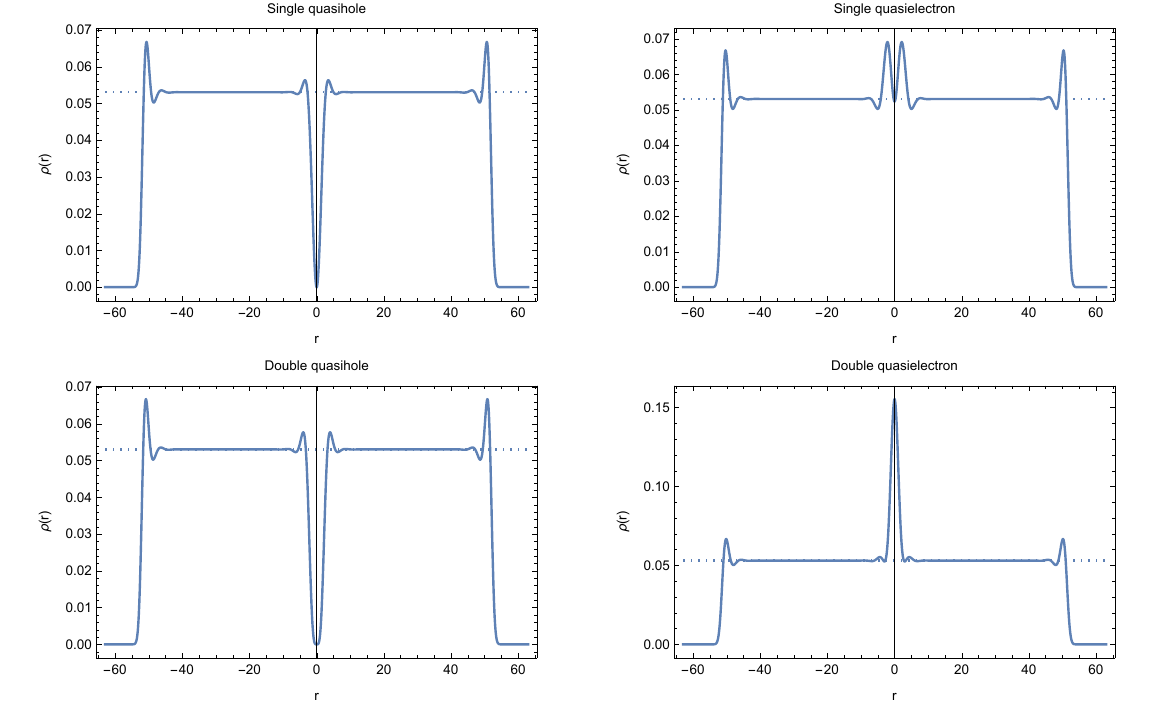}
\caption{%
The particle density profile $\rho(r)$ (in units of $1/\ell^2_B$) as a function of the radius (in units of $\ell_B$), for a single/double
quasihole (upper/lower left panel) and for a single/double Jain quasielectron (upper/lower right panel).
The dotted lines represent the background density $\rho_0 = \nu/(2\pi)$ with $\nu = 1/3$.
}
\label{fig:sm-densities}
\end{figure}

\begin{figure}
\includegraphics[width=\textwidth]{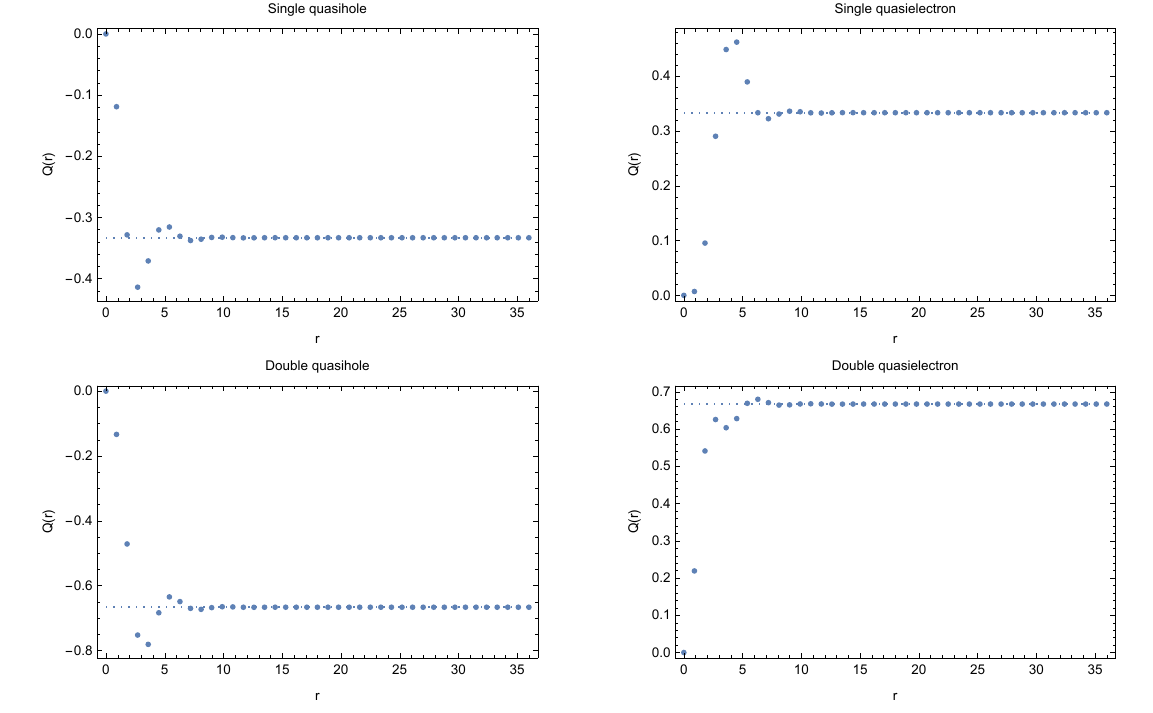}
\caption{
The excess charges $Q(r)$ (in units of $q$) as defined in the caption of Fig.~\ref{fig:ChargeAndDensity} for a single/double quasihole (upper/lower left panel) and for a single/double quasielectrons (upper/lower right panel).
The dotted lines represent the expected values $\pm \frac{1}{3}$ and $\pm \frac{2}{3}$.
}
\label{fig:sm-charges}
\end{figure}

\begin{figure}
\includegraphics[width=\textwidth]{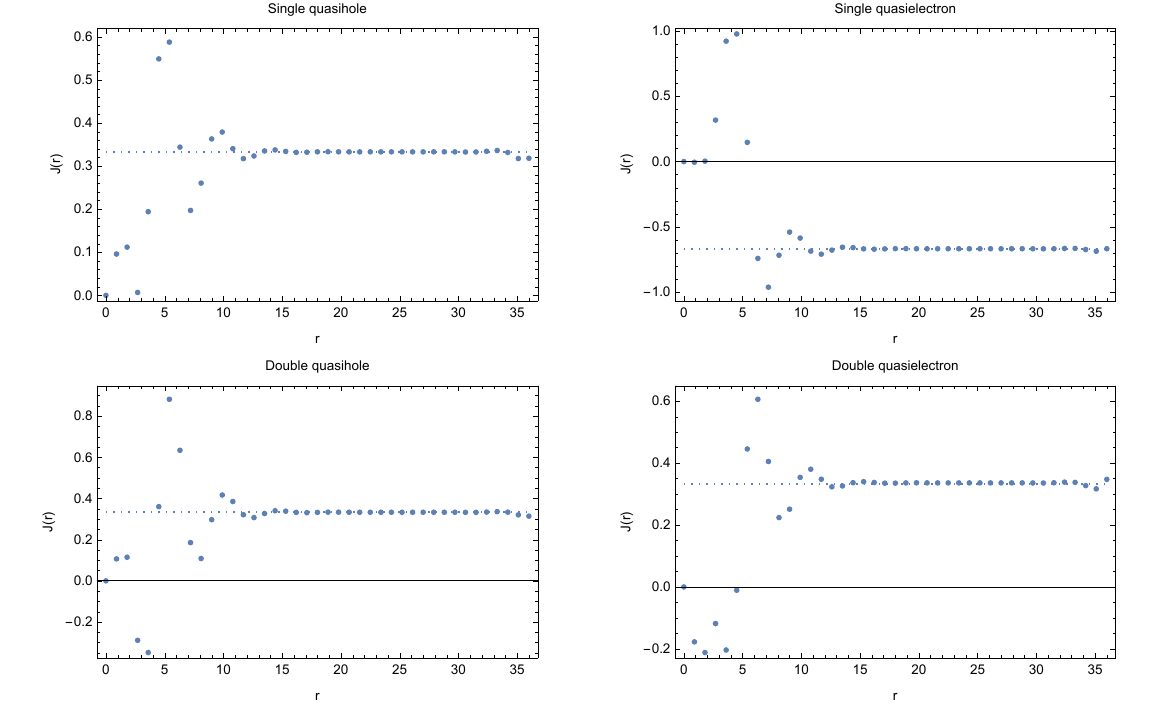}
\caption{
The spin $J(r)$ (in units of $\hbar$) for a single/double quasihole (upper/lower left panel) and for a single/double quasielectron (upper/lower right panel).
The dotted lines represent the expected values, namely $\frac{1}{3}$ for both the single and double quasihole, $-\frac{2}{3}$ for the single quasielectron and $\frac{1}{3}$ for the double quasielectron.
}
\label{fig:sm-spins}
\end{figure}

We first display the obtained density profiles in Fig.~\ref{fig:sm-densities}.
In Fig.~\ref{fig:sm-charges}, we display the excess charges $Q(r)$ (as defined above), as a function of the radial distance (in units of the magnetic length $\ell_B$).
Finally, in Fig.~\ref{fig:sm-spins}, we display the spin $J(r)$ as defined in the main text.

In calculating the integrals to obtain $Q(r)$ and $J(r)$, we assumed that the quasiparticles have cylindrical symmetry, and integrated only along the cylinder. The exception is the double QE. In that case, we performed the full 2-dimensional integral up to radius $r=\ell_B/2$. For $r>\ell_B/2$, we again assumed cylindrical symmetry for the double QE. The reason for doing this, is that in the case of the double QE, one needs a higher cutoff value $p_{\rm max}$ to obtain a cylindrically symmetric double QE.

The results we obtain using the MPS formulation of the quasihole and quasielectron states are fully consistent with the results obtained using Monte Carlo for the disk geometry. We should note, however, that the double quasielectron we simulate using the MPS formulation, has different short distance properties in comparison to the double Jain quasielectron studied in the main text.
It has been noted before in the literature, see for instance~\cite{Jeon_2003b, Hansson_2009b}, that there are different ways to create double quasielectron excitations.
The double quasielectron studied in the MPS formulation, has a charge profile that is more concentrated, leading to a higher value of the spin, in comparison to the double Jain quasielectron studied in the main text.
Because the difference between the two spins is an (even) integer, the results are nevertheless fully compatible with one another, the spin-statistics relation is satisfied in both cases.

\section{The non-abelian case}

In this section, we provide some details concerning the non-abelian case in general, and for the Moore-Read state~\cite{Moore_1991} in particular.
For non-abelian quantum Hall states, fusing two quasiparticles generically can lead to more than one different outcome.
This means that in the case of several quasiparticles, the ground state is typically degenerate.
However, for our purposes of obtaining a spin-statistics relation, we can focus on the case with two quasiparticles, for which the ground state is unique \footnote{Here, we make the assumption that the fusion rules for the quasiparticles do not have a fusion multiplicity, which is generically the case.}. 
Therefore, even the non-abelian case satisfies the assumptions we made in our derivation of the spin-statistics relation in the main text, in particular that the ground state is non-degenerate.
The non-abelian nature manifests itself via the possibility that fusion of two quasiparticles can lead to different results.
For each of these results, we have a different spin-statistics relation, that takes the same form as stated in the main text,
\begin{equation}
\kappa_{ab,c} = - J_{c} + J_a + J_b \pmod 1
\ ,
\end{equation}
where $c$ denotes the particular fusion outcome of fusing $a$ with $b$. 

\subsection{The Moore-Read case}

In this section, we explain how the spin-statistics relation works for the Moore-Read state \cite{Moore_1991}. This is confirmed by our numerical results.
We write the filling fraction of the state as $\nu = \frac{1}{q}$, where $q$ is even in the fermionic case.
The Moore-Read state is defined in terms of a chiral boson field $\varphi$ and the fields of the Ising conformal field theory (see f.i.~\cite{byb}).
This means that we should label the quasiholes by their Ising sector (i.e., $\id$, $\sigma$ or $\psi$), and their charge.
The smallest charge quasihole has the labels $\left(\sigma,\frac{1}{2q}\right)$.
Because the fusion of two $\sigma$ fields has two possible outcomes, $\sigma\times\sigma = 1+ \psi$, the fusion of two quasiholes also leads to two possible results.
In particular, we have (the charge label is additive, as is the case for the Laughlin state)
\begin{equation}
\left(\sigma,\frac{1}{2q}\right) \times \left(\sigma,\frac{1}{2q}\right) = \left(\id,\frac{1}{q}\right) + \left(\psi,\frac{1}{q}\right) \ .
\end{equation}
The first possible outcome $\left(\id,\frac{1}{q}\right)$ is the quasihole one obtains by piercing the sample with an additional flux, i.e., the ``ordinary" Laughlin quasihole.
The second possbile outcome ``contains" an additional neutral fermionic mode $\psi$.

From the explicit conformal field theory construction \cite{Moore_1991} (see also \cite{Bonderson_2011}) one obtains the statistical parameters of the double exchange of two charge $\frac{1}{2q}$ quasiholes, for both possible fusion outcomes.
For clarity, we drop the charge label when referring to the braid parameter $\kappa$. 
In particular, one finds $\kappa_{\sigma\sigma,\id} = \frac{1}{4q} - \frac{1}{8}$ (when the quasiholes fuse to $(\id,\frac{1}{q})$, the ordinary Laughlin quasihole).
For the other fusion channel, one has $\kappa_{\sigma\sigma,\psi} = \frac{1}{4q} + \frac{3}{8}$, that is, one has
$\kappa_{\sigma\sigma,\psi} = \kappa_{\sigma\sigma,\id} + h_\psi$, where $h_\psi = \frac{1}{2}$ is the scaling dimension of the neutral fermion.

We know, on theoretical grounds, the spin of the Laughlin quasihole in the Moore-Read state, that is, we know $J_{(\id,1/q)}$.
We can then first make a prediction for the spin of an elementary quasihole, $J_{(\sigma,1/(2q))}$, using the spin-statistics relation.
Finally, using $J_{(\sigma,1/(2q))}$, we can obtain the spin for the quasihole of type $\left(\psi,\frac{1}{q}\right)$, i.e. $J_{(\psi,1/q)}$.
In the following subsection, we provide numerical results, that confirm the values of the spin in these three cases.

Generically, the spin of a Laughlin quasihole is given by $J_{(\id,1/q)} = -\frac{1}{2q} + \frac{\mathcal{S}}{2q}$,
where $\mathcal{S}$ is the shift of the state, as can for instance be computed from the assumption of a rigid shift of the droplet's boundary~\cite{Comparin_2022}.
For the Moore-Read state, the shift is given by $\mathcal{S} = q+1$, which results in
\begin{equation}
J_{(\id,1/q)} = \frac{1}{2}.
\end{equation}
We find that the spin of the Laughlin quasihole in the Moore-Read state does not depend on the filling fraction.

By making use of $J_{(\id,1/q)} = \frac{1}{2}$, $\kappa_{\sigma\sigma,\id} = \frac{1}{4q} - \frac{1}{8}$, and the spin-statistics relation, we find
\begin{equation}
	J_{(\sigma,1/(2q))} = \frac{1}{2}( \kappa_{\sigma\sigma,\id} + J_{(\id,1/q)} ) = \frac{1}{8q}+\frac{3}{16}.
\end{equation}
For $q=1$, this results in $J_{(\sigma,1/2)} = \frac{5}{16}$, while for $q=2$, we have $J_{(\sigma,1/4)} = \frac{1}{4}$.

With the values of $J_{(\sigma,1/(2q))} = \frac{1}{8q}+\frac{3}{16}$ and $\kappa_{\sigma\sigma,\psi} = \frac{1}{4q} + \frac{3}{8}$ at hand,
we can now obtain $J_{(\psi,1/q)}$.
From the spin-statistics relation, we obtain
\begin{equation}
	J_{(\psi,1/q)} = 2  J_{(\sigma,1/(2q))} - \kappa_{\sigma\sigma,\psi} = 0 .
\end{equation}
Again, this value is independent of $q$.
In the following subsection, we numerically confirm the values of the spin obtained here.

\begin{figure}
	\centering
	\begin{minipage}{0.48\textwidth}
		\centering
		\includegraphics[width=1\textwidth]{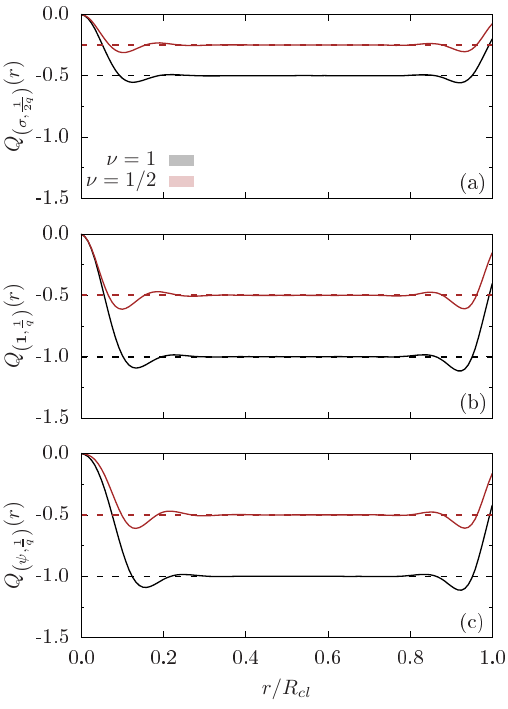} 
		\caption{Comparison of the quasihole charges $Q(r) = \int_0^r \left(\rho_{qp}(r')-\rho_{MR}(r')\right) r' dr'$, where $\rho_{qp}(r)$ is the QH density and $\rho_{MR}(r)$ the background Moore-Read density, for the different Moore-Read quasiholes: (a) the $\left(\sigma,\frac{1}{2q}\right)$, (b) the $\left(\id,\frac{1}{q}\right)$ and (c) the $\left(\psi,\frac{1}{q}\right)$, for the bosonic filling $\nu=1$ (corresponding to $q=1$) and the fermionic $\nu=\frac{1}{2}$ ($q=2$). 
		$R_{\rm cl}= \sqrt{2 N / \nu}$ is the classical radius of the droplet. 	\label{fig:MooreReadCharges}}
	\end{minipage}\hfill
	\begin{minipage}{0.48\textwidth}
		\centering
		\includegraphics[width=1\textwidth]{sm_MR_spins} 
		\caption{Comparison of the quasihole spins $J(r) = \int_0^r \left({r'}^2/2-1\right) \left(\rho_{qp}(r')-\rho_{MR}(r')\right) r' dr'$, where $\rho_{qp}(r)$ is the QH density and $\rho_{MR}(r)$ the background Moore-Read density, for the different Moore-Read quasiholes: (a) the $\left(\sigma,\frac{1}{2q}\right)$, (b) the $\left(\id,\frac{1}{q}\right)$ and (c) the $\left(\psi,\frac{1}{q}\right)$, for the bosonic filling $\nu=1$ (corresponding to $q=1$) and the fermionic $\nu=\frac{1}{2}$ ($q=2$). 
		$R_{\rm cl}= \sqrt{2 N / \nu}$ is the classical radius of the droplet. \label{fig:MooreReadSpins}}
	\end{minipage}
\end{figure}

\subsection{The Moore-Read case, numerical results}
In this Subsection we describe how we employed a Monte-Carlo sampling of the Moore-Read wavefunction in the presence of the different quasiholes $\left(\sigma,\frac{1}{2q}\right)$, $\left(\id,\frac{1}{q}\right)$ and $\left(\psi,\frac{1}{q}\right)$ in order to characterize their charges and spins.
Since the technique is analogous to that described in Section~\ref{Section:MCSampling}, here we just briefly describe the wavefunctions we considered.


We here always consider $N$ to be even because it is numerically simpler.
The ``Laughlin" quasihole $\left(\id,\frac{1}{q}\right)$ can be obtained by adiabatically piercing the system with a flux at position $\eta$, resulting in
\begin{equation}
	\label{eq:LaughlinQuasihole}
	\Psi_{\left(\id,\frac{1}{q}\right)}(\eta)=\prod_i^N (z_i-\eta)\,\text{Pf}\left(\frac{1}{z_i-z_j}\right)\,\prod_{i<j}^N(z_i-z_j)^q\,\exp\left(-\frac{1}{4}\sum_i |z_i|^2\right).
\end{equation}

The ``sigma" quasiholes $\left(\sigma,\frac{1}{2q}\right)$ are instead defined by ``splitting" a Laughlin quasihole making use of the properties of the Pfaffian factor
\begin{equation}
	\label{eq:SigmaQuasiholes}
	\Psi(\eta_1,\eta_2)=\text{Pf}\left(\frac{(z_i-\eta_1)(z_j-\eta_2)+(i \rightleftarrows j)}{z_i-z_j}\right)\,\prod_{i<j}^N(z_i-z_j)^q\,\exp\left(-\frac{1}{4}\sum_i |z_i|^2\right).
\end{equation}
From the numerical point of view, it is useful to maximize the distance between the quasiholes and the boundary of the system; the optimal solution is to place a single $\left(\sigma,\frac{1}{2q}\right)$ quasihole at $\eta=0$ -- the system's centre -- and send the other at spatial infinity. This will in general modify the properties of the boundary, but not those of the quasihole at the centre of the system.
We obtain
\begin{equation}
	\label{eq:SigmaQuasihole}
	\Psi_{\left(\sigma,\frac{1}{2q}\right)}(\eta=0)=\text{Pf}\left(\frac{z_i+z_j}{z_i-z_j}\right)\,\prod_{i<j}^N(z_i-z_j)^q\,\exp\left(-\frac{1}{4}\sum_i |z_i|^2\right).
\end{equation}

Finally, we introduce a quasihole $\left(\psi,\frac{1}{q}\right)$ by inspecting the four-$\left(\sigma,\frac{1}{2q}\right)$ quasiholes wavefunction.
Introducing the four quasi-hole ``building-block"
\begin{equation}
	\Psi_{(ab)(cd)}=\text{Pf}\left(\frac{(z_i-\eta_a)(z_i-\eta_b)(z_j-\eta_c)(z_j-\eta_d)+(i\rightleftarrows j)}{z_i-z_j}\right)\,\prod_{i<j}^N(z_i-z_j)^q\,\exp\left(-\frac{1}{4}\sum_i |z_i|^2\right)
\end{equation}
it is possible to define two degenerate four-quasihole states for suitable short-ranged Hamiltonians with pinning potentials~\cite{Nayak_1996}
\begin{equation}
	\begin{cases}
		\Psi_0=\prod_{\mu<\nu}^4\eta_{\mu\nu}^{\frac{1}{4q}-\frac{1}{8}} \frac{(\eta_{13}\eta_{24})^{\frac{1}{4}}}{\sqrt{1+\sqrt{1-x}}}\left(\Psi_{(13)(24)}+\sqrt{1-x}\,\Psi_{(14)(23)}\right)
		\\
		\Psi_1=\prod_{\mu<\nu}^4\eta_{\mu\nu}^{\frac{1}{4q}-\frac{1}{8}} \frac{(\eta_{13}\eta_{24})^{\frac{1}{4}}}{\sqrt{1-\sqrt{1-x}}}\left(\Psi_{(13)(24)}-\sqrt{1-x}\,\Psi_{(14)(23)}\right)		
	\end{cases}
\end{equation}
where $\eta_{\mu\nu}=\eta_\mu-\eta_\nu$ and $x=\frac{\eta_{12}\eta_{34}}{\eta_{13}\eta_{24}}$; these states are orthonormal~\cite{Bonderson_2011}.

By taking the appropriate limit of $\Psi_1$, we obtain
\begin{equation}
	\label{eq:PsiQuasihole}
	\Psi_{\left(\psi,\frac{1}{q}\right)}(\eta=0)\propto
	\text{Pf}\left(\frac{z_i^2+z_j^2}{z_i-z_j}\right)\,\prod_{i<j}^N(z_i-z_j)^q\,\exp\left(-\frac{1}{4}\sum_i |z_i|^2\right).
\end{equation}

In Fig.~\ref{fig:MooreReadCharges} and Fig.~\ref{fig:MooreReadSpins} we exhibit Monte Carlo results for the charge $Q(r) = \int_0^r \left(\rho_{qp}(r')-\rho_{L}(r')\right) r' dr'$ and spin Eq.~\eqref{eq:sm_spin_def} of the different Moore-Read quasiholes $\left(\sigma,\frac{1}{2q}\right)$ Eq.~\eqref{eq:SigmaQuasihole}, $\left(\id,\frac{1}{q}\right)$ Eq.~\eqref{eq:LaughlinQuasihole} and $\left(\psi,\frac{1}{q}\right)$ Eq.~\eqref{eq:PsiQuasihole}. 

It was not possible to run the Monte Carlo sampling on a GPU because of the larger space resources required to store the matrix $A_{ij}=\frac{1}{z_i-z_j}$; this indeed reflects on the larger errorbars for the Moore-Read spins in Fig.~\ref{fig:MooreReadSpins} as compared to the Jain's and Laughlin's quasielectron spins displayed in Fig. (2) of the main Article.
Nonetheless, the results for both the charge and spin are in agreement with the expectations, thereby proving the validity of our proof.



\end{document}